\documentclass[10pt]{article}
\usepackage{epsfig, rotate, multicol}
\begin{document}

\title{Medium effects in strange quark matter \\ and strange stars\footnote{Work supported 
by BMBF, GSI Darmstadt, and DFG}}
\author{K. Schertler, C. Greiner, and M.H. Thoma\footnote{Heisenberg fellow}\\
Institut f\"ur Theoretische Physik, Universit\"at Giessen,\\
35392 Giessen, Germany}

\date{November, 1996}
\maketitle

\begin{abstract}
We investigate the properties of strange quark matter at zero temperature 
including medium effects. The quarks are considered as quasiparticles which acquire an 
effective mass generated by the interaction with the other quarks of the 
dense system. The effective quark masses are derived from the zero momentum limit of the 
dispersion relations following from an effective quark propagator obtained from resumming
one-loop self energy diagrams in the hard dense loop approximation. This leads 
to a thermodynamic selfconsistent description of strange quark matter as 
an ideal Fermi gas of quasiparticles.
Within this approach we find that medium effects reduce 
the overall binding energy with respect to $^{56}F\!e$ of strange quark matter.
For realistic values
of the strong coupling constant strange quark matter is not
absolutely stable. The application to pure
strange quark matter stars shows that medium effects have, nevertheless, no impact 
on the mass-radius relation of the stars. However, a phase transition to hadronic matter at 
the surface of the stars becomes more likely.
\end{abstract}




\begin{multicols}{2}[\section{Introduction}]
Strange quark matter (SQM) or strangelets are thought to be confined
(bulk) objects containing a large number of delocalized
quarks $(u...u\, , \, d...d\, , \, s...s)$, so-called multiquark droplets.
Multiquark states consisting only of $u$- and $d$-quarks must have a mass
larger than ordinary nuclei, otherwise normal nuclei would
certainly be unstable, which, of course, is not the case.
However, the situation is different for
SQM, which would contain approximately the same
amount of $u$-, $d$- and $s$-quarks.
It has been suggested as a possible
absolutely stable or metastable
phase of nuclear matter \cite{SQM}, which might be realized in form  of small
droplets (stranglets) or in the interior of superdense stars \cite{MH}.
The proposal that hypothetical SQM
at zero temperature and in $\beta$-equilibrium might be absolutely
stable has stimulated substantial activity. Such a scenario
would be of fundamental importance, as, for example, it could
explain a large fraction of the non-observable mass of the universe
in certain cosmological models, and could modify substantially
our understanding of the structure and stability of neutron stars \cite{MH}.
The reason for the possible stability lies in introducing a third flavor degree of freedom,
the strangeness, where the mass of the strange quarks is considerably smaller
than the Fermi energy of the quarks, thus lowering the total mass
per unit baryon number of the system.
The equation of state of SQM has been described as a non-interacting
gas of quarks at zero temperature,
taking into account the phenomenological bag constant. Also quark interactions within lowest order perturbative QCD have been considered \cite{FM,CK}.
First studies in the context of the MIT-bag model predicted that
sufficiently heavy strangelets might be absolutely stable \cite{FJ}, i.e.,
that the energy per baryon might be lower than the one of $^{56}F\!e$.
On the other hand, for smaller strangelets it is also conceivable that the mass per baryon is
lower than the mass of the strange $\Lambda $-baryon, but larger than the nucleon
mass. The droplet is then in a metastable state, i.e., it cannot decay
(strongly) into nonstrange hadronic matter \cite{CK}.
Metastable strangelets relax somewhat the
stringent conditions on the choice of parameters of the bag model
required by an absolutely stable state \cite{FJ}, namely small
bag constants $B_0^{1/4} \stackrel{< }{\sim } 150$ MeV.
According to this picture, SQM should appear as a nearly neutral and massive
state because the number of strange quarks is nearly equal to the number
of massless up or down quarks and so the strange quarks neutralize that
hypothetical form of nuclear matter.

In condensed matter as well as nuclear physics medium effects play an 
important role. One of the most important medium effects are effective
masses generated by the interaction of the particles with the system.
In a gluon gas at high temperature the consideration of an effective
mass for the gluons within the ideal gas approximation
lead to an excellent description of the equation of state found in lattice 
calculations \cite{GS,PKPS}.
The effective gluon mass has been derived from the gluon self energy, 
containing the interaction of the gluons with the heat bath. Using the
high temperature approximation \cite{KW} or, equivalently, the hard
thermal loop (HTL) approximation \cite{BP}, which ensures a gauge invariant
treatment, a gluon self energy of the order $g^2T^2$ has been found.
Resumming this self energy leads to an effective quark propagator. The
dispersion relation of the gluons in the plasma follows from the pole of
this effective propagator. Due to the complicated momentum
dependence of the HTL gluon self energy the dispersion relation can be
obtained only numerically. Besides the branch belonging to 
transverse gluons, a plasmon branch shows up, describing the propagation 
of a longitudinal gluon in the plasma. The plasma frequency, defined 
as the zero momentum limit of the dispersion relation, is given by
$m_g^*=gT/\sqrt {3}$. Instead of using the complicated dispersion 
relation, the transverse gluons have been regarded as quasiparticles
with the plasma frequency as effective mass. The plasmons, on the other 
hand, have been neglected as they are suppressed for typical momenta
of the order of the temperature \cite{PIS}. Although the effective masses have been
calculated under the assumption of weak coupling ($g<1$), the resulting
equation of state is in good agreement with the one found on the lattice,
even close to the phase transition where the coupling constant is not
small. 

Here we want to apply the idea of an ideal gas of quasiparticles with
effective masses to the case of cold (strange) quark matter. Instead of gluons at
finite temperature we consider up, down, and strange quarks at zero 
temperature but finite chemical potential $\mu $. The effective quark 
mass is obtained from the zero momentum limit of the quark dispersion
relation, which follows from the hard dense loop (HDL) approximation
\cite{MA} of the quark self energy at finite chemical potential, where
the internal momenta of the one-loop self energy diagram  are assumed
to be hard, i.e. of the order $\mu $ or larger. This calculation will be
presented in the next section, where we focus in particular on the 
role of the current quark masses. In the third section we will discuss
the statistical mechanics of a system
of quasi-particles with density dependent masses.
The effects of these masses on the equation of state of quark matter
and a comparision with previous results without medium effects will be
the topic of section 4. Finally we will apply our results to quark stars
in section 5. 

\end{multicols}

\vspace{1cm}
\begin{multicols}{2}[\section{Medium-dependent effective quark masses}]

The aim of the present section  is the derivation of an effective quark mass
in an extended, dense system of quarks at zero temperature. This effective 
mass is defined as the zero momentum limit of the dispersion relation
following from the quark self energy, which can be written in the most
general form as \cite{BO}
\begin{equation} \label{self}
\Sigma =-a\> P_\mu \gamma ^\mu -b\> \gamma _0-c,
\end{equation}
where 
\begin{eqnarray} \label{abc}
a&=&\frac{1}{4p^2}\> [tr(P_\mu \gamma ^\mu \Sigma )-p_0\, tr(\gamma _0
\Sigma )], \\ \nonumber
b&=&\frac{1}{4p^2}\> [P^2\, tr(\gamma _0\Sigma )-p_0 tr(P_\mu \gamma^\mu
\Sigma )], \\ \nonumber
c&=&-\frac{1}{4}\> tr\Sigma, 
\end{eqnarray}
and $P^2=p_0^2-p^2$.

\begin{figure}[ht]
\centerline{{\epsfig{file=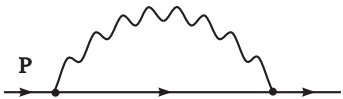,height=8cm}}}
\caption{One-loop quark self energy}
\label{oneloop}
\end{figure}

The dispersion relation is determined from the poles of the effective 
propagator
\begin{equation} \label{prop}
S^*(P)=\frac{1}{P_\mu \gamma ^\mu -m-\Sigma},
\end{equation}
leading to
\begin{equation} \label{disp}
[p_0(1+a)+b]^2=p^2(1+a)^2+(m-c)^2,
\end{equation}
where $m$ is the current quark mass.
The traces over $\Sigma $ are calculated in the hard dense loop (HDL)
approximation \cite{MA}, where only hard momenta of the order of the quark
chemical potential $\mu $ for the internal quark and gluon lines in the 
one-loop quark self energy of fig.\,\ref{oneloop} are used. As in the high temperature case 
\cite{BP} the vacuum contribution and the contribution from integration over
soft momenta are of higher order. This approximation is equivalent to the
leading term in the high density approximation analogously to the finite 
temperature case \cite{KW}. In this way we arrive at a gauge invariant 
expression for the self energy, complete to leading order in the coupling
constant $g$. Of course, due to the perturbative nature of this approximation,
the resulting  expressions are only valid for small coupling constants,
corresponding to a large chemical potential $\mu $, i.e. high density,
according to asymptotic freedom. However, as shown in the case of a hot
gluon gas, the equation of state using effective masses calculated
perturbatively agrees very well with lattice results even at temperatures
where the coupling constant is not small \cite{PKPS}. (As a matter of fact, there
are indications that for large coupling constants the lowest order results
are better than higher order corrections \cite{PPC}.)

In the case of a vanishing current quark mass, as it can be assumed for
up and down quarks, the HDL quark self energy is given by \cite{KW,VT}
\begin{eqnarray} \label{HDL}
tr\Sigma &=& 0,\\ \nonumber
tr(P_\mu \gamma ^\mu \Sigma)&=& 4{m_q^*}^2, \\ \nonumber
tr(\gamma _0 \Sigma)&=&2{m_q^*}^2\> \frac{1}{p}\> \ln \frac{p_0+p}{p_0-p}, 
\end {eqnarray}
where 
\begin{equation} \label{defmq}
{m_q^*}^2=\frac{g^2\mu ^2}{6\pi ^2}.
\end{equation}
(In the finite temperature case the only
change is the replacement of ${m_q^*}^2$ by $g^2(T^2+\mu ^2/\pi ^2)/6$ 
\cite{VT}.)

Resumming this quark self energy leads to an effective propagator (\ref{prop}) which
is part of a systematic expansion (Braaten-Pisarski method) at finite
temperature and chemical potential \cite{BP,VT}. (For a review of this 
method as 
well as its applications to the quark-gluon plasma see \cite{QGP}.)

For non-vanishing current quark masses, e.g. for strange quarks with
$m_s \simeq 150$ MeV, we have two possibilities for incorporating the current
mass into the effective one. First the current mass could be hard, i.e. of the
order of $\mu $. In this case we have a second hard scale besides the chemical
potential. Then the HDL approximation for the quark self energy may not be
sufficient, as terms of the order of $g m_s$ in the quark self energy from
the integration over soft momenta or from the vacuum contribution may appear.
Also gauge invariance of the self energy is not longer guaranteed.
On the other hand, if we assume the current mass to be soft, $m_s \sim g\mu $,
it can be neglected in the calculation of the quark self energy, where only
hard momenta and energies are needed. The soft mass assumption might be
justified, even for small $g$ since $\mu \simeq 350$ MeV holds,
as we will see below. Hence we end up with the same quark self energy as in 
the massless case (\ref{HDL}) and (\ref{defmq}).

\begin{figure}[ht]
\centerline{\rotate[r]{\epsfig{file=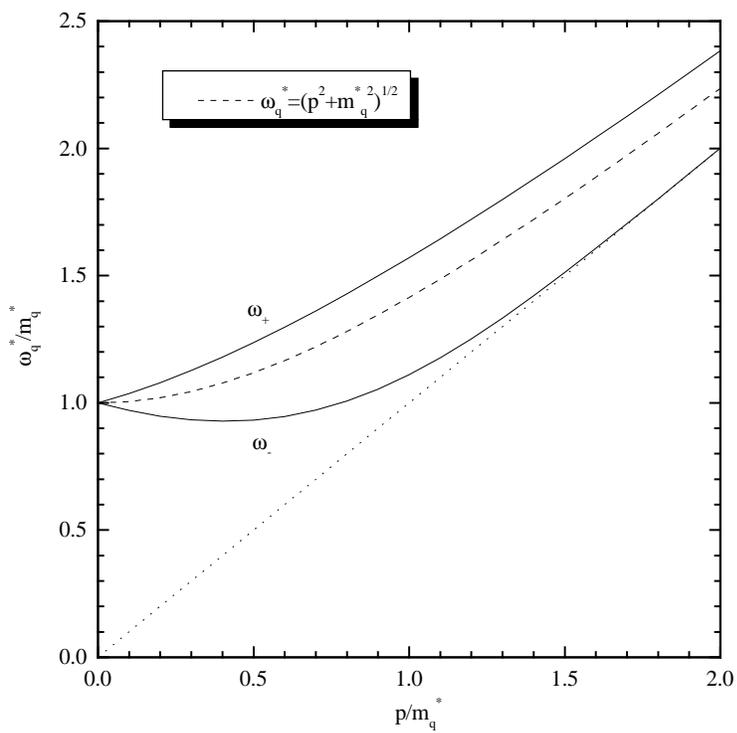,height=14.5cm}}}
\caption{Quark dispersion relations}
\label{dispers}
\end{figure}

Inserting (\ref{HDL}), (\ref{defmq}), 
and (\ref{abc}) into (\ref{disp}) in the 
massless case leads to the dispersion relation shown in fig.\,\ref{dispers}. The upper
branch $\omega_+$ corresponds to a collective quark mode, whereas the lower branch $\omega_-$
describes the propagation of a so-called plasmino \cite{BPY}, which is absent 
in the vacuum analogously to the plasmon solution for gluons. The effective 
mass at $p=0$ is given by $m_q^*$. In the case of a current mass of the order
of $m_s\sim g\mu $ we must not neglect this mass in (\ref{prop}) and (\ref{disp}) since
$m_s$ and $\Sigma $ are of the same order. In the zero momentum limit
we find from (\ref{disp}) the effective strange quark mass \cite{BO}
\begin{equation} \label{defms}
m_s^*=\frac{m_s}{2}+\sqrt {\frac{m_s^2}{4}+\frac{g^2\mu ^2}{6\pi ^2}}.
\end{equation}

For evaluating the equation of state for an ideal gas of collective quarks,
we neglect the plasmino as it is suppressed, i.e., the spectral
strength function (residue of
the pole of the effective proagator) vanishes exponentially for hard momenta of the order
of $\mu $ \cite{WEL} which dominate the equation of state. Furthermore we 
replace the dispersion relation of the collective quark branch by
\begin{equation} \label{dispsimp}
\omega^*_{q,s}(k) = \sqrt {k^2+{m_{q,s}^*}^2}.
\end{equation}
The simple dispersion relation $\omega^*_q$ is shown in fig.\,\ref{dispers}. It deviates 
from the exact one $\omega_+$ by less than 11\% over the entire momentum range. 
The deviation of $\omega^*_s$ from the exact solution can be shown to be even smaller.

Since there are no lattice calculation at 
finite density to which we could compare our results, the simple dispersion relation
(\ref{dispsimp}) is sufficient for our purpose, describing the role of medium effects
in SQM at least qualitatively. Therefore we neglect further improvements of the 
quasiparticle treatment introduced in the case of a hot gluon gas \cite{PKPS} as 
the use of an effective number of degrees of freedom, of a temperature dependent coupling 
constant, and of an additional constant in front of the effective mass in the dispersion
relation for fitting the high momentum part of the exact dispersion relation.  

Finally we will discuss the value of $g$ to be taken here. In principle it should be 
determined as a $\mu $-dependent running coupling constant from the renormalisation
group equation at finite density. However, so far there are no clear results at finite 
temperature or density \cite{RUN}. Assuming the zero density result 
\begin{equation} \label{runcoup}
\alpha _s (Q)=\frac {g^2}{4\pi }= \frac {12\pi }{(33-2n_f)\ln (Q^2/\Lambda ^2)}
\end{equation}
and replacing $Q$ by the average three momentum in quark matter $\langle q\rangle =3\mu /4$
we find for three quark flavors ($n_f=3$), $\mu  =350$ MeV (see below), and the QCD
scale parameter $\Lambda = 200$ MeV a value of $g=5.7$. In calculations of SQM including
lowest order $\alpha _s$-corrections typical values of $g=5.3$
($\alpha_s=2.2$) have been used \cite{CK},
derived from comparing the MIT bag model with hadronic properties. In the approach
proposed here one expects that $g$ ought to be small as the effective mass is calculated
perturbatively. However, in the case of a gluon gas the use of the lowest order effective
mass provided an excellent agreement with lattice results even at the phase transition, where
$g=7.7$ was used \cite{PKPS}. As we will see below, an upper limit for the coupling constant
in our approach is given by $g=7.7$. As medium effects in quark matter cannot be related 
directly to $\alpha _s$-corrections fitted to hadronic properties and the zero density 
result of the running coupling constant cannot be trusted, we take $g$ as a free parameter ranging
from 0 to 5. 
\end{multicols}
\vspace{1cm}

\begin{multicols}{2}[\section{Statistical mechanics of systems with medium dependent effective quark masses}]

The underlying philosophy to do statistical mechanics with medium dependent effective quark masses is to treat an ensemble of quasiparticles as
a free, (in this case) degenerate Fermi gas of particles with a total energy
\begin{eqnarray}
\label{Hamiltonian}
\hat{H}_{eff} & = &
\sum_{i=1}^{d} \sum _{{\bf k}} \sqrt{{\bf k}^2+m_i^{*2}(\mu )}
 \ \hat{a}^{\dagger }_{{\bf k},i} \hat{a}_{{\bf k},i} \\
 & & +E^{*}(\mu ). \nonumber
\end{eqnarray}
Here $d$ denotes the degree of degeneracy (e.g. $d=6 n_f$ for $n_f$ flavors). 
Due to the $\mu$-dependence of $m^*(\mu )$ the function
$E^*(\mu )=B^*(\mu ) V$ defines a necessary energy counterterm
in order to maintain thermodynamic self-consistency
\cite{GorensteinYang1995}.
From this Hamiltonian the particle density $\rho$, energy density $\epsilon$ and
pressure $p$ at temperature $T=0$ of a Fermi gas of free quasiparticles are given by

\begin{eqnarray} \label{defrhogeneral}
\rho(\mu) & = & \frac{d}{2 \pi^2} \int_{k=0}^{k_F} dk \, k^2 \\
 & = &  \frac{d}{6 \pi^2} k_F^3, \nonumber
\end{eqnarray}

\begin{eqnarray} \label{defepsgeneral}
\epsilon(\mu) - B^*(\mu) & = & \frac{d}{2 \pi^2} \int\limits_{k=0}^{k_F} dk \, k^2 \omega^*(k) \\
 & = &  \frac{d}{16 \pi^2} \left[ \mu \, k_F \left( 2 \mu^2 - {m^*}^2 \right) \right. \nonumber\\
 & &  \left. - {m^*}^4 \ln {\left( \frac{k_F + \mu}{m^*}\right)} \right], \nonumber
\end{eqnarray}

\samepage{
\begin{eqnarray} \label{defpgeneral}
p(\mu) + B^*(\mu) & = & \frac{d}{2 \pi^2} \int\limits_{k=0}^{k_F} dk \, k^2 (\mu - \omega^*(k))  \\
 & = & \frac{d}{24 \pi^2} \left[ \mu \, k_F \left( \mu^2 - \frac{5}{2} {m^*}^2 \right) \right. \nonumber \\
  & & \left. +\frac{3}{2} {m^*}^4 \ln {\left( \frac{k_F + \mu}{m^*} \right)} \right]. \nonumber
\end{eqnarray}
}
Up to the additional function $B^*$, which can be regarded as a $\mu $-dependent bag constant,
these are the ideal Fermi gas formulas at temperature $T=0$ for quasiparticles of mass
$m^*$ and chemical potential $\mu$. The Fermi momentum is $k_F = (\mu^2 - {m^*}^2)^{1/2}$.
The function $B^*(\mu )$ is necessary to maintain thermodynamic self-consistency.
For example the fundamental thermodynamic relation between energy density and pressure

\begin{equation} \label{thermoeps}
\epsilon(\mu)  =  \mu \frac{d p(\mu)}{d \mu} -p(\mu)
\end{equation}
has to be fulfilled. So we should require additionally \cite{GorensteinYang1995}

\begin{equation}
\left( \frac{\partial p}{\partial m^*}\right)_\mu = 0.
\end{equation}
By using (\ref{defpgeneral}) we then find

\begin{equation} \label{defBstargeneral}
  \frac{d B^*}{d m^*}  =  -\frac{d}{4 \pi^2} \left[ m^* \mu \, k_F - {m^*}^3
    \ln {\left(\frac{k_F + \mu}{m^*}\right)} \right].
\end{equation}
To carry out the integration of (\ref{defBstargeneral}) we have to use the
explicit functional dependence $m^*(\mu)$ respectively $\mu = \mu(m^*)$ in the two cases of light- and strange quarks.

\end{multicols}


\vspace{1cm}

\begin{multicols}{2}[\subsection{EOS of massless quarks} \label{lightquarks}]

In the case of $m^*(\mu)= m^*_q(\mu)$ (\ref{defmq}) we integrate
(\ref{defBstargeneral}) and obtain

\begin{equation} \label{Bq}
  B^*_q(\mu_q) = -\frac{d_q}{16 \pi^2} \left[ \alpha^2 \beta - \alpha^4 \ln 
  {\left(\frac{\beta+1}{\alpha} \right)} \right] \mu_q^4 \, ,
\end{equation}
choosing $B^*_q(0)=0$. (Later we will add the phenomenological bag constant $B_0$.)
The $g$ dependent functions $\alpha(g)$ and $\beta(g)$ are given by
\begin{eqnarray} \label{defalpha}
	\alpha(g)& = & \frac{g}{\sqrt{6} \pi}, \\
	\beta(g) & = & \sqrt{1-\alpha^2(g)}. \label{defbeta}
\end{eqnarray}
After inserting (\ref{Bq}) into (\ref{defrhogeneral})-(\ref{defpgeneral}) and some manipulations using $m^*=\alpha(g) \mu$,
the equations of state then assume the simple form

\begin{equation} \label{defrhoq}
  \rho_q(\mu_q) = \frac{d_q}{6 \pi^2} \beta^3 \mu_q^3 ,
\end{equation}

\begin{equation} \label{defepsq}
  \epsilon_q(\mu_q) = \frac{d_q}{8 \pi^2} \beta^3 \mu_q^4 ,
\end{equation}

\begin{equation} \label{defpq}
  p_q(\mu_q) = \frac{d_q}{24 \pi^2} \beta^3 \mu_q^4,
\end{equation}
which obviously fulfill $\rho_q=\partial p_q/ \partial \mu_q$ and $\epsilon_q +
p_q = \mu_q \rho_q$. 
They are the well known equations for massless fermions up to the
factor $\beta^3$ (\mbox{$0<\beta(g)<1$}). Note that $\rho_q$, $\epsilon_q$ and
$p_q$ decreases with increasing coupling constant $g$. On the other hand the implicit 
form $p_q=p_q(\epsilon_q)$ of the EOS $p_q = \epsilon_q/3$ does not reveal any $g$ 
dependence, while the EOS in the form $p_q=p_q(\rho_q) \propto \rho^{4/3}/\beta(g)$ 
slightly depends on g. The EOS in the form $p_q=p_q(\epsilon_q)$ (with respect to 
modifications due to massive $s$-quarks and electrons) will be important for the 
investigations of SQM-stars in section \ref{stars}. From (\ref{defbeta}) we observe 
that $g$ must not exceed the value $g=\sqrt{6}\pi \simeq 7.7$.

To get an idea of the order of magnitude of $B_q^*$ (which is negative), fig.\,\ref{bqmu} shows $|B_q^*|^{1/4}$ versus $\mu_q$ for various values of $g$. If $\mu \approx 350$ MeV holds (see below) $|B_q^*|^{1/4}$ is of the order of $100$ MeV. Using (\ref{Bq}) and (\ref{defepsq}) this corresponds to $B_q^*/\epsilon_q < 10 \%$ for $g<5$. 

\end{multicols}


\begin{multicols}{2}[\subsection{EOS of massive quarks} \label{strangequarks}]

Taking $m^*(\mu)$ in (\ref{defBstargeneral}) to be the effective strange quark
mass $m^*_s(\mu_s)$ given by (\ref{defms}), the integration leads to
($d_s=6$)

\begin{eqnarray} \label{Bstars}
  B^*_s(\mu_s) & = &-\frac{d_s}{16 \pi^2} \left[ \frac{\sqrt{(m^*_s-m_s)(\beta^2 m^*_s
        -m_s)}}{24 \alpha^2\beta^4} \right.  \nonumber \\ 
 & & \times \left. \sum_{n=0}^3 a_n m_s^{3-n} {m_s^*}^{n} \right.  \\ 
 & & + \frac{5 \alpha^4 -12 \alpha^2 +8}{16 \beta^5} m_s^4 \nonumber \\
 & & \times \ln {\left(\frac{\beta\sqrt{m^*_s-m_s}+\sqrt{\beta^2 m^*_s-m_s}}
     {\alpha^2 m_s}\right)} \nonumber \\ & &
\left. - {m^*_s}^4 \ln {\left( \frac{k_F + \mu}{m^*_s} \right)} \right], \nonumber
\end{eqnarray}
where
\begin{eqnarray*}
  	a_0(g) & = & -15 \alpha^4 +26 \alpha^2 -8, \\ 
	a_1(g) & = & -10 \alpha^4 +18 \alpha^2 -8, \nonumber \\
	a_2(g) & = & -8 \alpha^4 +16 \alpha^2 -8, \nonumber \\
	a_3(g) & = & 24 \alpha^4 -48 \alpha^2 +24.
  \nonumber
\end{eqnarray*}
From requiring a positive Fermi momentum, $k_F=(\mu ^2-{m^*}^2)^{1/2}\geq 0$, we find
$\mu _s\geq m_s^*$, from which $\mu _s\geq m_s\beta ^{-2}$ follows. This constraint is 
equivalent to $\beta ^2m_s^*\geq m_s$ needed in (\ref{Bstars}). 
At $\mu_s = m_s \beta^{-2}$ the integration constant is chosen in such a way that $B^*_s$ vanishes. Utilizing
(\ref{Bstars}) in connection with (\ref{defrhogeneral})-(\ref{defpgeneral})
(using $m^*=m^*_s$)
leads to the equations of state $\rho_s(\mu_s)$, $\epsilon_s(\mu_s)$ and $p_s(\mu_s)$ 
for strange quarks.

\end{multicols}

\vspace{1cm}
\section{Medium-effects in strange quark matter}

In the following sections we investigate in detail within our model the equation of state and the properties of SQM.

\begin{multicols}{2}[\subsection{Strange quark matter consisting of massless quarks}]

It is illustrative to consider first the case of SQM consisting of massless 
quarks ($m_{u,d,s}=0$).
In section \ref{lightquarks} we derived the equations of state
(\ref{defrhoq})-(\ref{defpq}) for light quarks with the effective
quark mass (\ref{defmq}). 
We saw that $\rho_q$, $\epsilon_q$ and $p_q$ decreases with
increasing coupling constant $g$. Now we want to investigate how this affects the energy per
baryon ($E/A$) of massless SQM

\begin{eqnarray}
  \left(\frac{E}{A}\right)(\mu_q) & = & \frac{\epsilon_q(\mu_q) + B_0}{\rho_B(\mu_q)} \\
 & = & 3 \left(\frac{3}{4} \mu_q + \frac{6 \pi^2}{d_q \beta^3 \mu_q^3} B_0 \right). \nonumber
\end{eqnarray}
Here $B_0$ denotes the usual bag-constant which corresponds to the energy difference between the
"false", perturbative vacuum inside the bag and the "true" vacuum at the
outside, and which, in return, confines the quarks. The degree of degeneracy is given by $d_q=6 n_f=18$ and
$\rho_B(\mu_q)=\rho_q(\mu_q)/3$ is the baryon density. The minimum energy
with respect to $\mu_q$ is found to be

\begin{equation} \label{EdurchAminud}
  \left(\frac{E}{A}\right)_{min} = 3 \mu_{min},
\end{equation}
where
\begin{equation}
  \mu_{min} = \left( \frac{24 \pi^2}{d_q \beta^3} B_0 \right)^{1/4}
\end{equation}
is the chemical potential which minimizes $E/A$. The optimum chemical potential $\mu _{min}$
ranges from 276 MeV for $g=0$ and $B_0^{1/4}= 145$ MeV to 374 MeV for $g=5$ and $B_0^{1/4}=160$
MeV. Minimizing the energy is here equivalent to the equilibrium condition $p=p_q-B_0=0$ between the Fermi pressure inside the bag and the pressure of the nonpertubative QCD vacuum outside the 
bag given by $B_0$. This leads to an electrically neutral system with an equal number of $u$, 
$d$ and $s$-quarks.
Eq. (\ref{EdurchAminud}) shows that in the case of massless quark matter the medium effects, we
take into account, lead to an increase of the energy per baryon by the
factor 
\begin{displaymath} \mu_{min} \propto \beta^{-3/4}= \left( 1-\frac{g^2}{6 \pi^2} \right)^{-3/8}. \end{displaymath}
For instance assuming $B_0^{1/4}=145$ MeV ($B_0=58$ MeV/fm$^3$) this increases
$(E/A)_{min}$ from about $829$ MeV ($g=0$) to $881$ MeV ($g=3$, $\alpha_s=0.72$). 
That is to say, owing to medium effects the binding energy per baryon of the system with
respect to $^{56}F\!e$ is reduced from about $100$ MeV to $50$ MeV. The more
massive constituents raise the overall energy scale.

At $p=0$ ($\mu=\mu_{min}$) we find using (\ref{defrhoq})-(\ref{defpq}) that 
\[\rho_q(\mu_{min}) \propto \beta^3 \mu_{min}^3 \propto \beta^{3/4} \]
decreases while 
\[\epsilon_q(\mu_{min}) \propto \beta^3 \mu_{min}^4 \propto \beta^3 \beta^{-3}=\mbox{const}.\]
is independent of $g$.

\end{multicols}


\begin{multicols}{2}[\subsection{Strange quark matter}]

\begin{figure}[ht]
\centerline{\rotate[r]{\epsfig{file=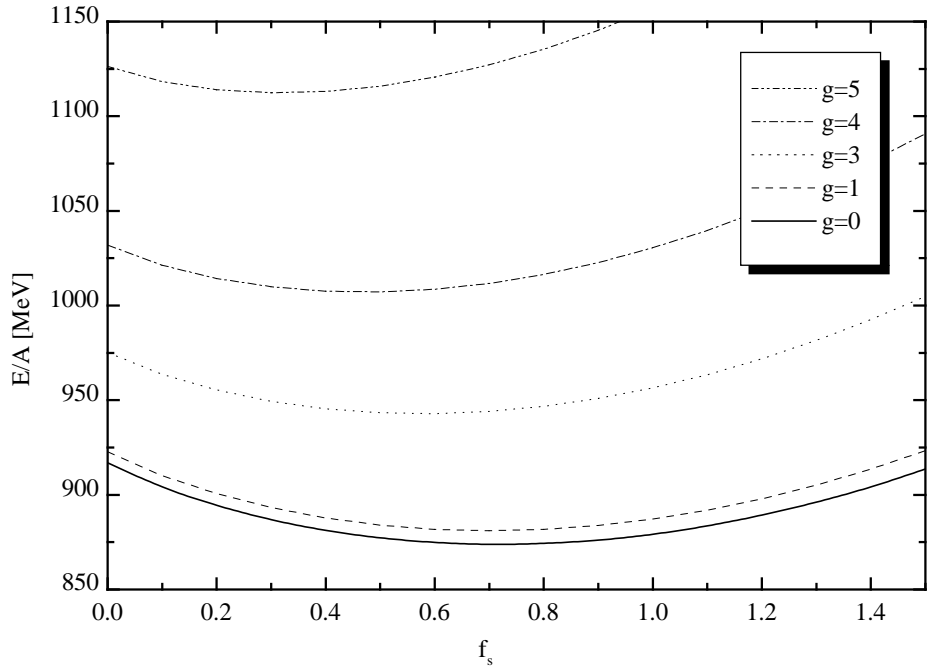,height=14.5cm}}}
\caption{Energy per baryon versus the strangeness fraction at $B_0^{1/4}=145$ MeV and $m_s=150$ MeV}
\label{eafs}
\end{figure}

Now we use the equations of state derived in section \ref{lightquarks} and \ref{strangequarks} to
calculate the energy per baryon of strange quark matter ($d_q=12$, $d_s=6$) assuming finite $s$-quark masses:

\begin{equation} \label{EdurchAs}
  \left(\frac{E}{A}\right)(\mu_q, \mu_s) = \frac{\epsilon_q(\mu_q) +\epsilon_s(\mu_s)  +
    B_0}{\rho_B(\mu_q, \mu_s)}.
\end{equation}
The baryon density is now given by

\begin{equation}
\rho_B(\mu_q, \mu_s) = \frac{\rho_q(\mu_q)+\rho_s(\mu_s)}{3}. 
\end{equation}
As usual we define the strangeness fraction $f_s$ to be

\begin{equation} 
f_s(\mu_q,\mu_s)=\frac{\rho_s(\mu_s)}{\rho_B(\mu_q,\mu_s)},
\end{equation}
yielding the number of $s$-quarks per baryon. Fig.\,\ref{eafs} shows the minimum energy per baryon $E/A$ (i.e. $p=p_u+p_d+p_s-B_0=0$) for a given strangeness content $f_s$ and various values of $g$. The bag constant is assumed to be $B_0^{1/4}=145$ MeV. The current $s$-quark mass is $m_s=150$ MeV. One sees that $E/A$ increases with increasing $g$. At $g=0$ the {\em absolute} minimum is achieved for a strangeness fraction of $f_s \approx 0.7$. The absolute minimum corresponds to a system where all flavors assume the same Fermi energy (i.e. $\mu_q = \mu_s$).

\begin{figure}[ht]
\centerline{\rotate[r]{\epsfig{file=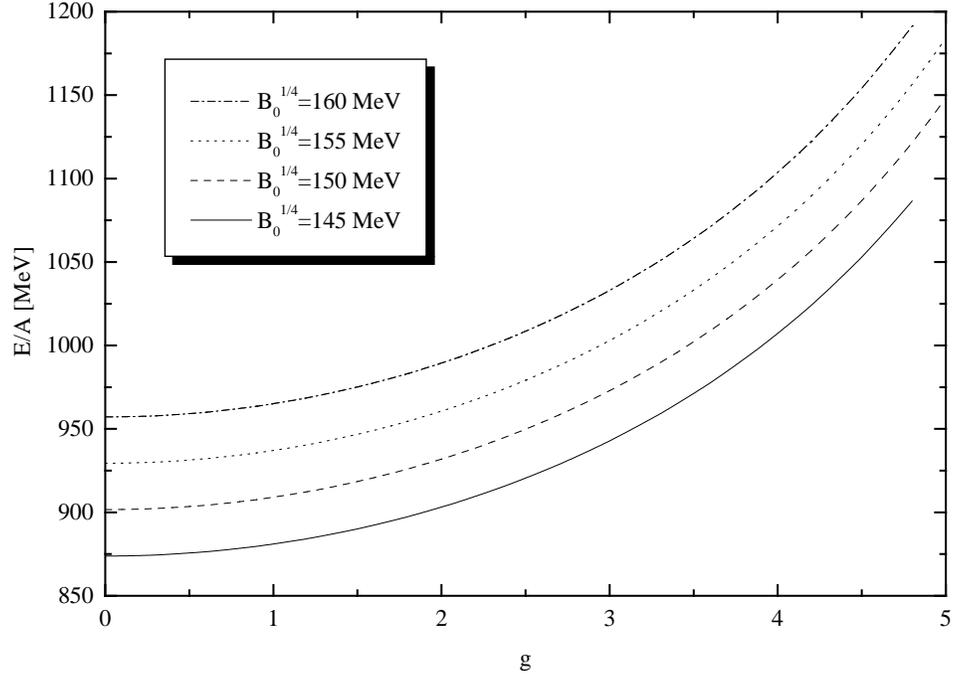,height=14.5cm}}}
\caption{Energy per baryon versus the coupling constant at $m_s=150$ MeV}
\label{eag}
\end{figure}

\begin{figure}[ht]
\centerline{\rotate[r]{\epsfig{file=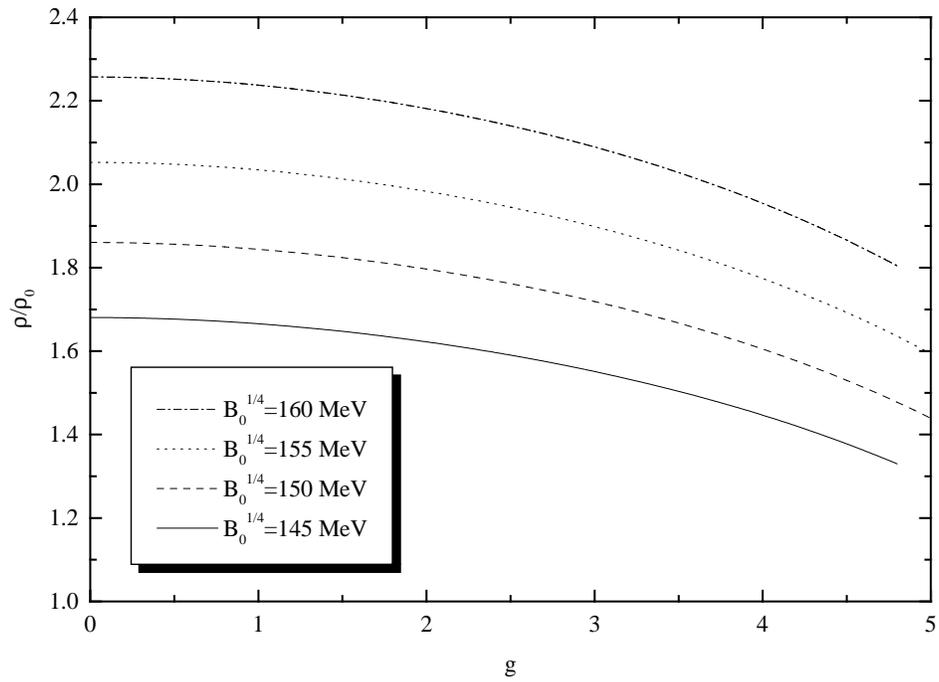,height=14.5cm}}}
\caption{Baryon density versus the coupling constant at $m_s=150$ MeV, $\rho_0=0.17 fm^{-3}$}
\label{rhog}
\end{figure}

\begin{figure}[ht]
\centerline{\rotate[r]{\epsfig{file=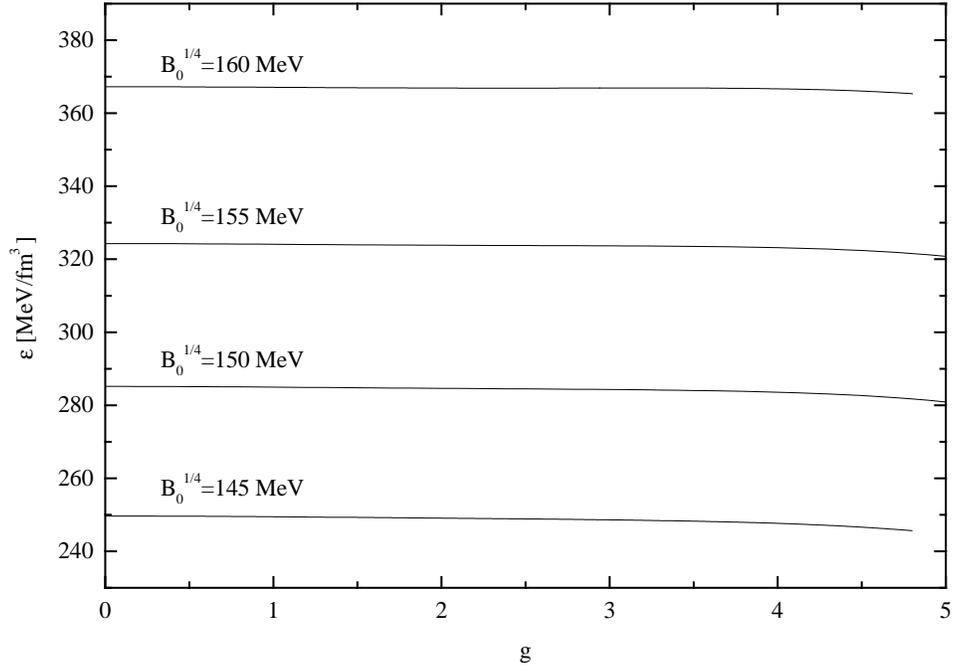,height=14.5cm}}}
\caption{Energy density versus the coupling constant at $m_s=150$ MeV}
\label{epsg}
\end{figure}

\begin{figure}[ht]
\centerline{\rotate[r]{\epsfig{file=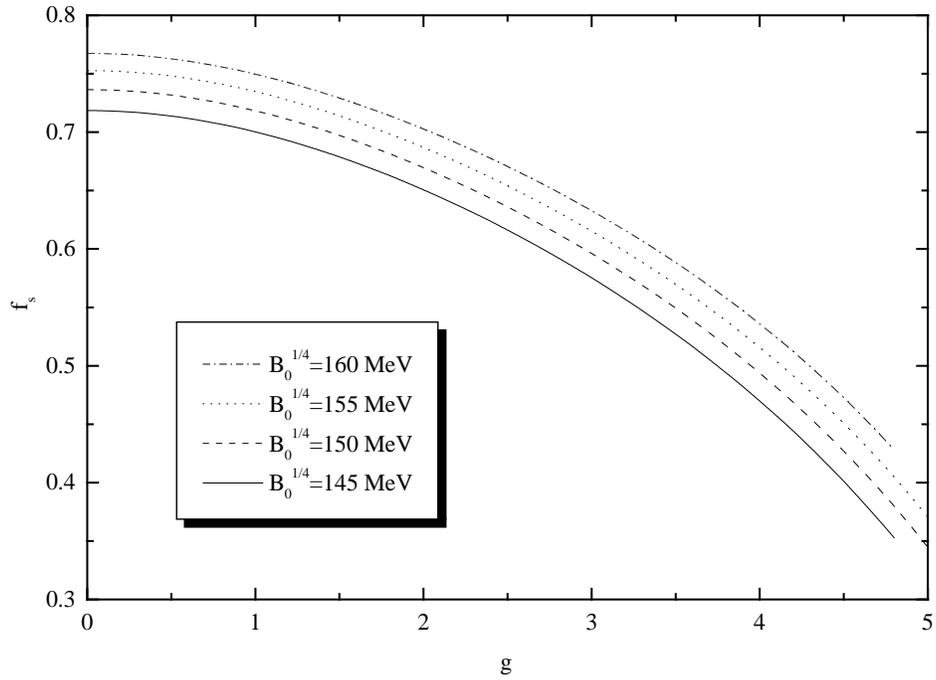,height=14.5cm}}}
\caption{Strangeness fraction versus the coupling constant at $m_s=150$ MeV}
\label{fsg}
\end{figure}

In the following we focus on the properties of SQM in its {\em absolute} minimum. Fig.\,\ref{eag} shows this minimal energy versus g. Assuming $m_s=150$ MeV and $B_0^{1/4}=145$ MeV the minimal energy increases from $874$ MeV ($g=0$) to $943$ MeV ($g=3$). Therefore, for the chosen parameters SQM is not bound with respect to $^{56}F\!e$.
As already seen in the massless case ($m_s=0$) the baryon density at $p=0$ decreases with 
increasing $g$ (fig.\,\ref{rhog}) while the energy density $\epsilon$ stays approximately 
constant (fig.\,\ref{epsg}).
The strangeness fraction $f_s$ is found to decrease with increasing $g$ (fig.\,\ref{fsg}) which 
is in contrast to the massless calculation where $f_s=1$ due to flavor symmetry. Furthermore, 
this result is also in contrast to the effect of $\alpha_s$-corrections on $f_s$ which will be
discussed in the next section. The decreasing strangeness fraction does not seem to be an 
artifact of the special form of the effective mass $m^*_s$ given by (\ref{defms}). For instance 
the same behaviour holds for a "hypothetical" mass of the form

\[m^*_s = m_s + \frac{g \mu}{\sqrt{6} \pi}. \]

So, in this approach medium effects lead to less dense and less bound strange quark matter 
with a lowered number of strange quarks per baryon!

\end{multicols}


\begin{multicols}{2}[\subsection{Medium effects versus bag constant}]

\begin{figure}[ht]
\centerline{\rotate[r]{\epsfig{file=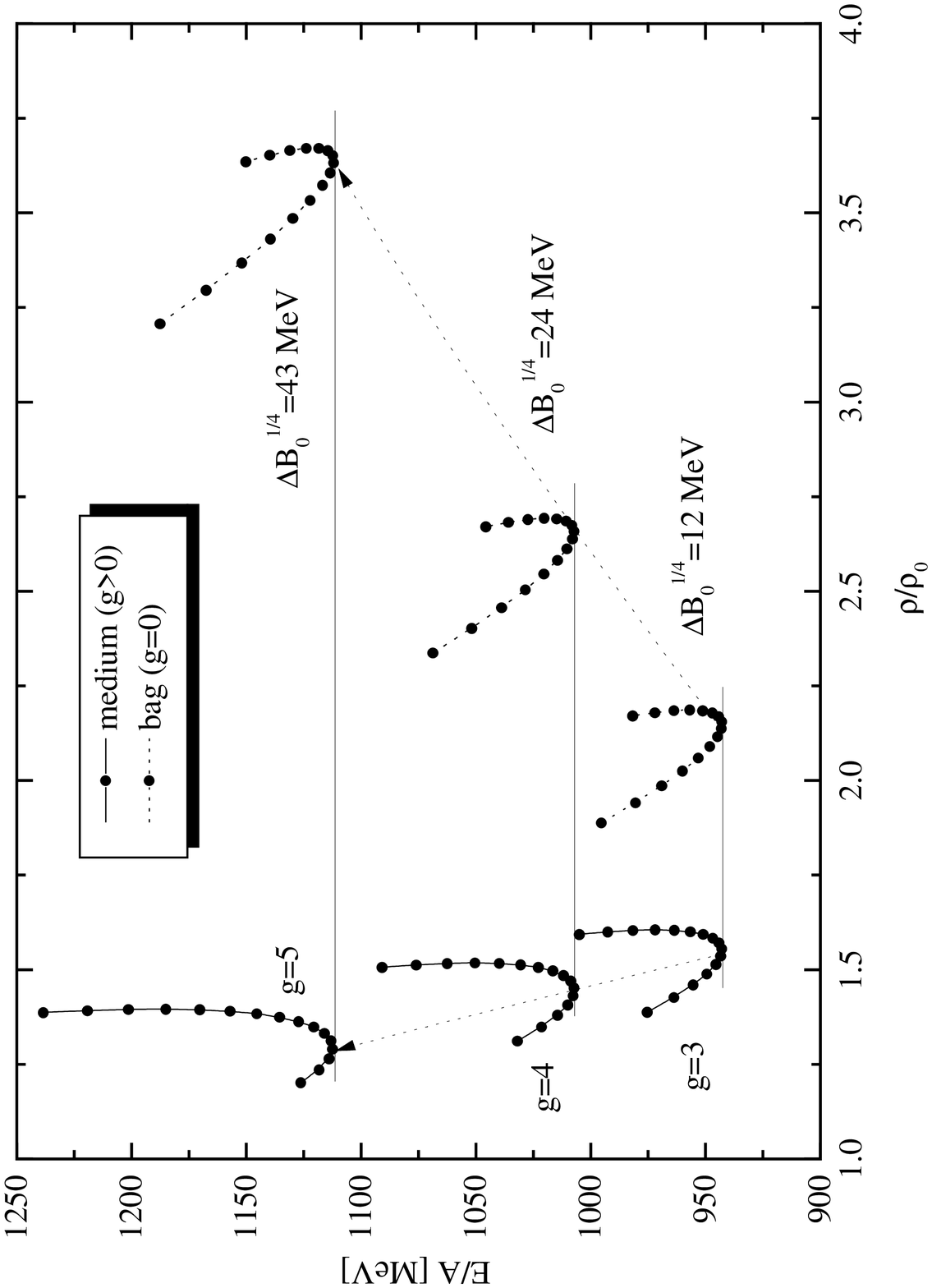,height=14.5cm}}}
\caption{Comparison of $\rho$ in the "medium" approach and the "bag" approach at 
$m_s=150$ MeV and $B_0^{1/4}=145$ MeV. The points denote the increase of $f_s$ from 
$f_s=0$ (top left) to $f_s=1.5$ (top right) in steps of 0.1; $\rho_0=0.17 fm^{-3}$}
\label{earhocmp}
\end{figure}

\begin{figure}[ht]
\centerline{\rotate[r]{\epsfig{file=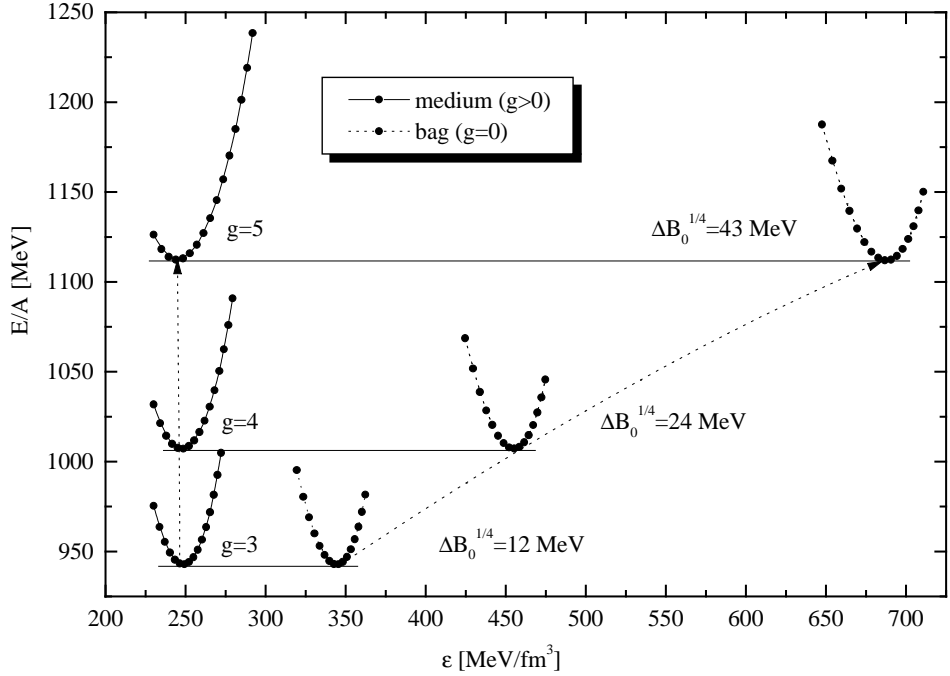,height=14.5cm}}}
\caption{Comparison of $\epsilon$ in the "medium" approach and the "bag" approach at 
$m_s=150$ MeV and $B_0^{1/4}=145$ MeV. The points denote the increase of $f_s$ from 
$f_s=0$ (top left) to $f_s=1.5$ (top right) in steps of 0.1}
\label{eaepscmp}
\end{figure}

As shown above, describing SQM with medium effects using effective quark masses leads to the 
mass formula (\ref{EdurchAs}) depending on the additional parameter $g$.
From a practical point of view, introducing the new parameter $g$ gives rise to the question 
whether the effect of changing $g$, say the effect of the medium, can already be described by
a variation of the original parameters that enter the model, namely $B_0$ and $m_s$. The 
bag constant might be suitable for this purpose because of the uncertainty of its 
value and its strong influence on the behavior of SQM (figs.\,\ref{eag}-\ref{fsg}).
Therefore we will briefly discuss whether a change in the bag constant ($B_0 \rightarrow 
B_0+\Delta B_0$) at $g=0$ is able to approximately recover changes of the properties 
($E/A$, $\rho$, $\epsilon$) due to medium effects parameterized by $g$.
As a criterion to compare the "medium" approach (where $B_0$ is unchanged and $g>0$) 
with the "bag" approach (changing $B_0$ at $g=0$) we require that both should lead to 
the same energy per baryon at its absolute minimum, i.e.
\begin{equation} \label{criterion}
(E/A)_{med}(B_0, g>0) \stackrel{!}{=} (E/A)_{bag}(B_0+\Delta B_0, g=0).
\end{equation}
Here "med" and "bag" stands for the "medium" and the "bag" approach, respectively.
Under this condition we calculate $\rho_{med}$, $\epsilon_{med}$ for different values 
of $g$ and compare it to the associated $\rho_{bag}$, $\epsilon_{bag}$ at a $\Delta B_0$ 
so that (\ref{criterion}) holds. Fig.\,\ref{earhocmp} and fig.\,\ref{eaepscmp} show the minimal energy for various $f_s$ (see figure caption) versus $\rho$ and $\epsilon$. 
One sees that both approaches lead to a different physical behavior.
Whereas $\rho_{med}$, $\epsilon_{med}$ decreases (or stays constant) with increasing $g$, 
one sees that $\rho_{bag}$, $\epsilon_{bag}$ strongly increases. The same different behavior appears 
for $f_s$. From this we can conclude that it is not possible to describe the influence
of medium effects just by altering the bag constant. For completeness, we should mention that
similar arguments hold for the variation of $m_s$ instead of $B_0$.

Furthermore, the EOS of SQM has been computed including $\alpha_s$-corrections, i.e. one-gluon
exchange \cite{FM, Greiner}. Increasing $\alpha _s$ there leads to an increase of $E/A$ as in our 
approach. However, at the same time $f_s$ increases in contrast to our case. Hence, we conclude
that the influence of the effective quark masses on SQM cannot be simulated by taking
into account $\alpha _s$-corrections or changing $B_0$ or $m_s$.

\end{multicols}
\vspace{0.5cm}
\section{Application to strange quark matter stars} \label{stars}

Originally SQM in bulk was thought to exist
only in the interior of neutron stars where the pressure is 
high enough that the neutron matter melts into its quark substructure
\cite{Col75}. At least in the cores of neutron stars, where the density rises up to the order of $10$ times normal nuclear density, it is not very likely that matter consists of individual hadrons. It thus still might be that SQM exists in neutron stars.
For an overview of the structure of neutron stars, see Ref. \cite{WeberGlendenning1996}. 

\begin{multicols}{2}

In this section we investigate neutron stars under the assumption that they {\em entirely} consist 
of cold electrically charge neutral SQM (SQM stars) in equilibrium with respect to the weak 
interaction. This requires the inclusion of electrons into the model.
We focus on the change of the properties of SQM-stars ($p(\epsilon)$, $M$, $R$, $E/A$) 
due to medium effects in the spirit of the previous sections. Here $M$ and $R$ denotes the mass and radius of the star.

The thermodynamic of electrons at $T=0$ is given by 
(\ref{defrhogeneral})-(\ref{defpgeneral}) taking $d=2$ and $m^*$=0
\begin{eqnarray}
\rho_e & = & \frac{\mu_e^3}{3 \pi^2}, \\
\epsilon_e & = & \frac{\mu_e^4}{4 \pi^2}, \\
p_e & = & \frac{\mu_e^4}{12 \pi^2}.
\end{eqnarray}
Now, we are dealing with four chemical potentials ($\mu_u$, $\mu_d$, $\mu_s$, $\mu_e$) 
which are related by the chemical equilibrium between the quark flavors and the leptons. 
The basic weak reactions are given by

\begin{eqnarray}
d & \longrightarrow & u+e^- +\bar{\nu}_{e^-} \\
s & \longrightarrow & u+e^- +\bar{\nu}_{e^-}.
\end{eqnarray}
The equilibration of flavors is provided by \cite{EQU}

\begin{equation}
s + u \longrightarrow d + u.
\end{equation}
Hence, the four chemical potentials are reduced to two independent ones

\begin{equation}
\mu \equiv \mu_s = \mu_d \qquad \mbox{and} \qquad \mu_u = \mu - \mu_e.
\end{equation}
Finally, the condition of electrically charge neutrality

\begin{equation} 2/3 \rho_u -1/3 (\rho_d + \rho_s) -\rho_e = 0 \end{equation}
just leaves one independent chemical potential, say $\mu$. Therefore, the EOS
can be written as a function of $\mu $ only:

\begin{eqnarray}
\rho_B& = & (\rho_u+\rho_d+\rho_s)/3, \\
\epsilon & = & \epsilon_u + \epsilon_d +\epsilon_s +\epsilon_e +B_0 \label{rhostar}, \\
p & = & p_u +p_d +p_s +p_e -B_0, \label{pstar}
\end{eqnarray}
where $\rho_{u,d}$, $\epsilon_{u,d}$ and $p_{u,d}$ are calculated from 
(\ref{defrhoq})-(\ref{defpq}) with $d_q=6$ and individual chemical potentials $\mu_u$, $\mu_d$.

\begin{figure}[ht]
\centerline{\rotate[r]{\epsfig{file=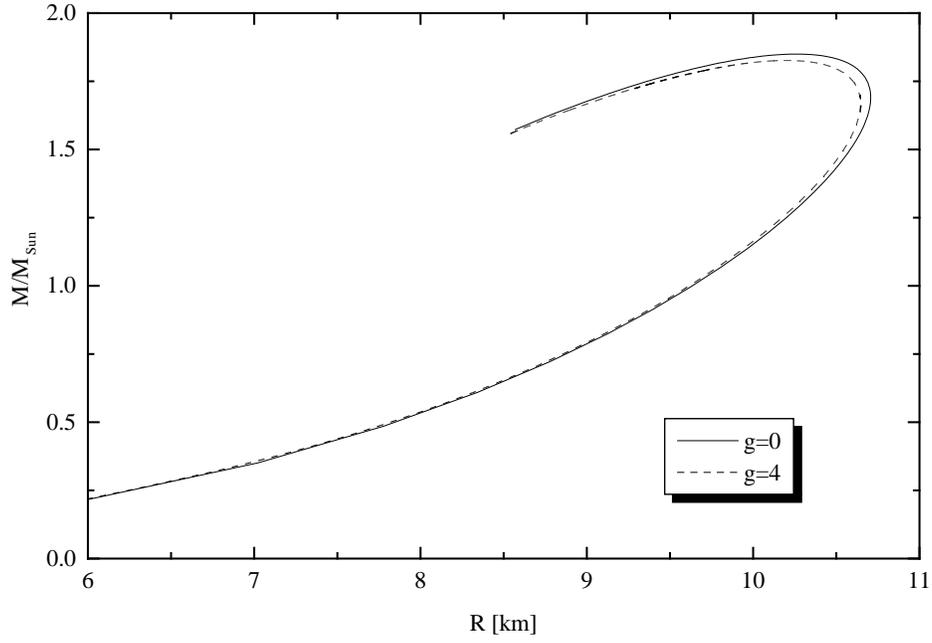,height=14.5cm}}}
\caption{Mass (in units of solar mass) versus radius of a SQM star, $m_s=150$ MeV, $B_0^{1/4}=145$ MeV}
\label{mr}
\end{figure}

One of the aim of this section is to determine whether there are significant changes 
in mass $M$ and radius $R$ of SQM stars due to medium effects. To do so, we assume a 
cold, static, spherical star. It is described by the solutions of the 
Oppenheimer-Volkoff-Tolman (OVT) equations of hydrostatic equilibrium.
They follow from general relativity \cite{OVT} which one has to apply due to highly 
concentrated matter and therefore curved space-time. The solutions of the OVT equations 
are the pressure $p(r)$ and energy density $\epsilon(r)$ at a given radius $r$ inside the star. 
The radius $R$ of the star is determined by the condition $p(R)=0$ while the mass is given by

\begin{equation} M=\int_{r=0}^R 4 \pi r^2 \epsilon(r) dr. \end{equation}
The OVT equation can be solved specifying the central energy density $\epsilon_c=\epsilon(r=0)$ 
and the EOS in the form $p=p(\epsilon)$. The numerical calculation of $p=p(\epsilon)$ from 
(\ref{rhostar}), (\ref{pstar}) shows that there is no noteworthy influence by a change of $g$. 
The EOS gets slightly softer for high values of the coupling constant (e.g. $g=4$) but there is 
no remarkable impact on $R$ and $M$ (see fig.\,\ref{mr}).

\begin{figure}[ht]
\centerline{\rotate[r]{\epsfig{file=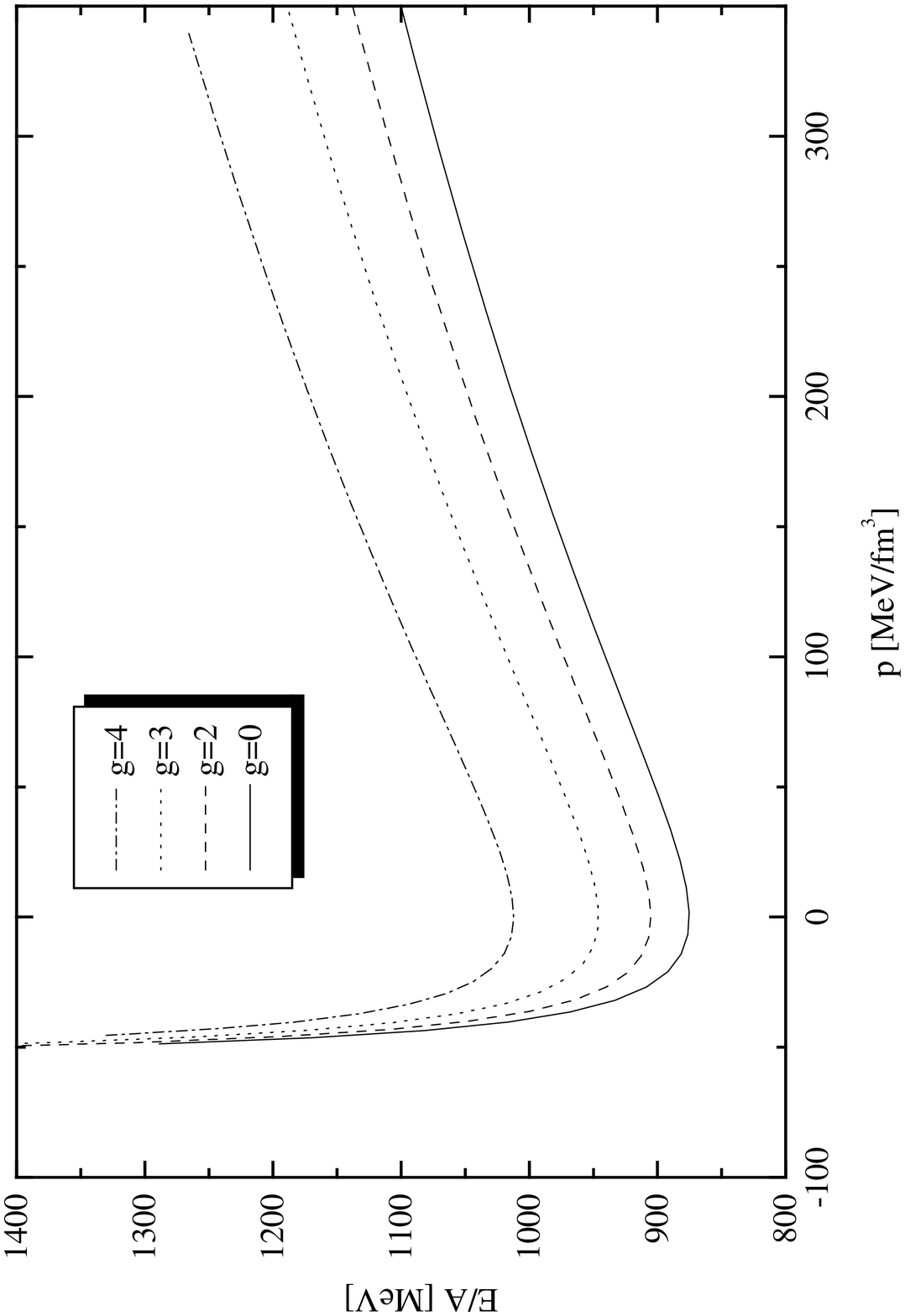,height=14.5cm}}}
\caption{$E/A$ versus $p$ in a SQM star, $m_s=150$ MeV and $B_0^{1/4}=145$ MeV}
\label{eap}
\end{figure}

\begin{figure}[ht]
\centerline{\rotate[r]{\epsfig{file=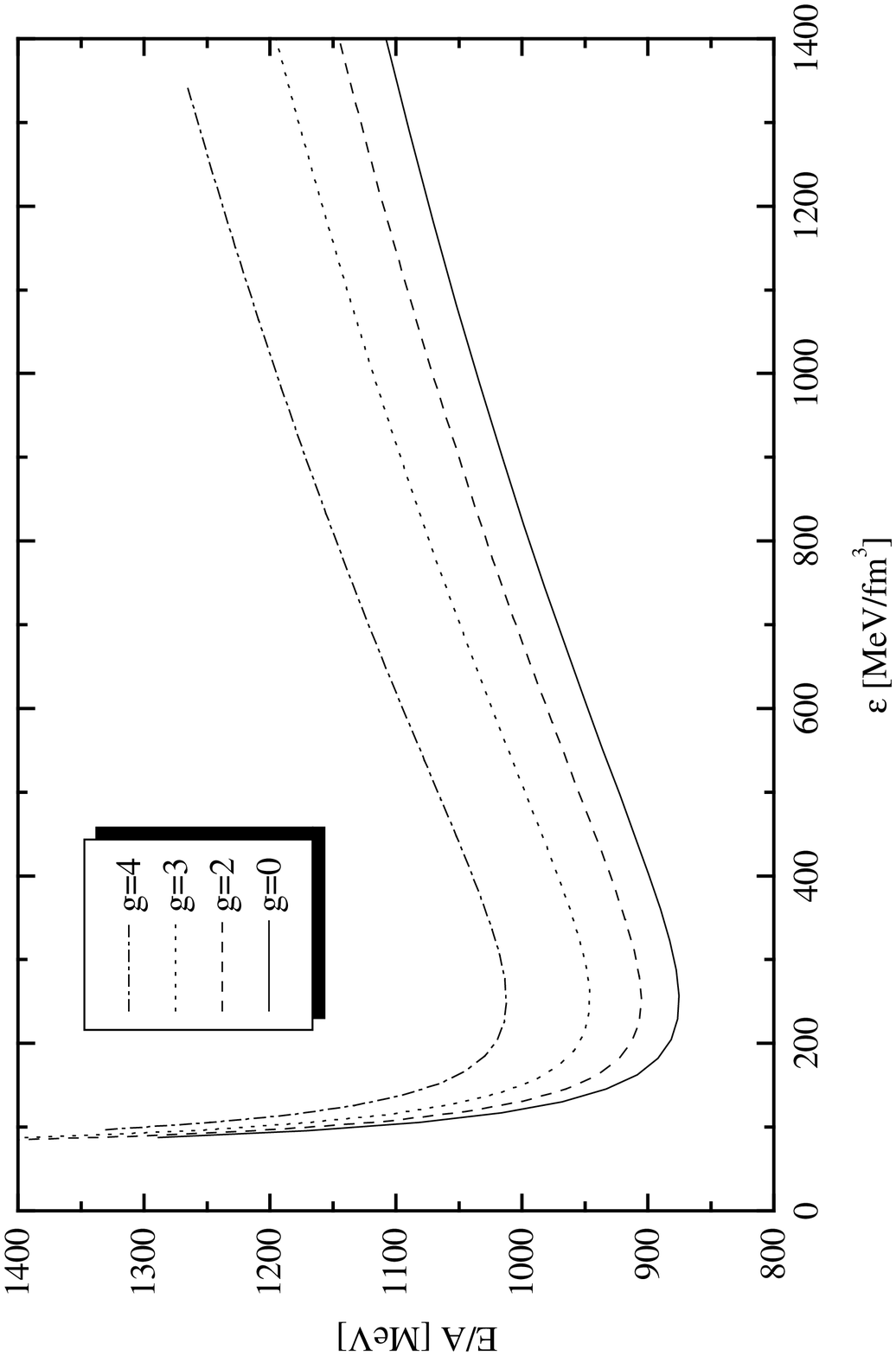,height=14.5cm}}}
\caption{$E/A$ versus $\epsilon$ in a SQM star, $m_s=150$ MeV and $B_0^{1/4}=145$ MeV}
\label{eaeps}
\end{figure}

\begin{figure}[ht]
\centerline{\rotate[r]{\epsfig{file=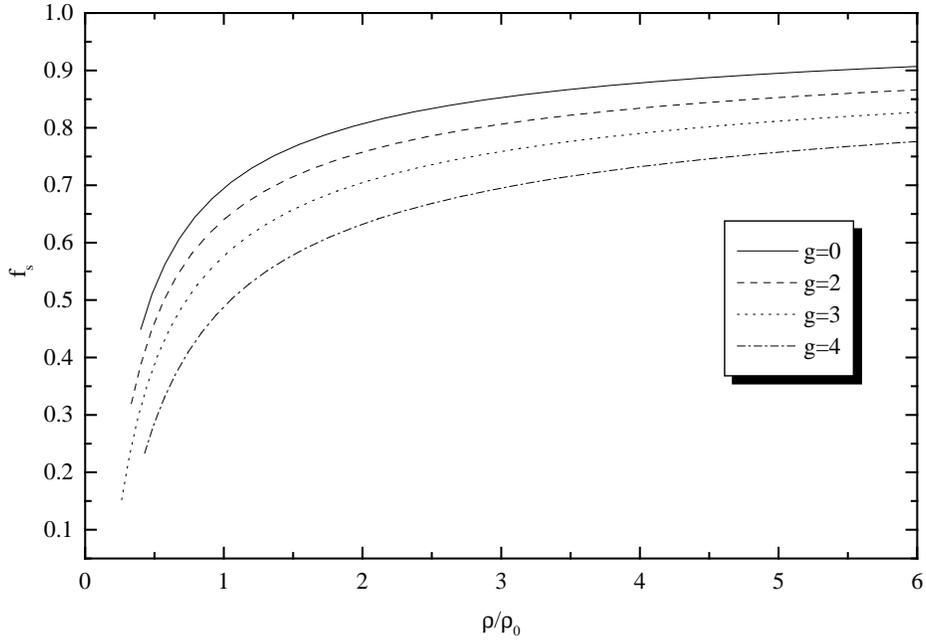,height=14.5cm}}}
\caption{Strangeness fraction $f_s$ versus $\rho/\rho_0$ in a SQM star, $m_s=150$ MeV, $B_0^{1/4}=145$ MeV and $\rho_0=0.17 fm^{-3}$}
\label{fsrho}
\end{figure}

Note, that this result is only valid under the assumption of a pure SQM star. Although there 
is no change in mass and radius of the star, there is, nevertheless, an increase of the energy 
$E/A$ at a given radius inside the star due to medium effects.
This increases the possibility of a phase transition to
hadronic matter and therefore the 
possibility of a nuclear crust on the surface of the SQM core.
Figs.\,\ref{eap} and \ref{eaeps} 
show this behavior as a function of $p$ and $\epsilon$, respectively.
In fig.\,\ref{eap} the pressure inside the star decreases from a given central pressure 
$p_c=p(\epsilon_c)$ in the center of the star to $p=0$ on the surface where $E/A$ is minimal. 
One finds that the energy $E/A$ is increased in the entire star. Furthermore, fig.\,\ref{fsrho} 
shows the change of the strangeness fraction versus the baryon density for different values of $g$.

We conclude the discussion of strange stars with the result that mass and radius of pure SQM 
stars are not changed by medium effects while, considering hybrid stars, mass and radius of 
the star might be indirectly influenced by medium effects due to a change of the  crust formation.

\end{multicols}

\begin{multicols}{2}[\section{Conclusions}]

In the present work we proposed an improved description of SQM by taking medium effects
into account. We considered the quarks as quasiparticles, which acquire an effective mass
generated by the interaction with other quarks of the dense system. The effective masses
are derived from the zero momentum limit of the dispersion relations
following from the quark self energy in the hard dense loop approximation. We regard the 
strong coupling constant $g$ contained in the effective mass as a free parameter.
Although the quark self energy has been calculated perturbatively we extrapolate our
final results up to $g=5$. The validity of this approach is supported
by the success of an equivalent method applied to a hot gluon gas \cite{PKPS}.

In the case of quark matter consisting of quarks with zero current mass the density 
(\ref{defrhoq}), the energy density (\ref{defepsq}), and the pressure (\ref{defpq}) 
are simply changed by a factor $\beta^3(g)$ (\ref{defbeta}), which decreases
with increasing $g$ starting from $\beta =1$ at $g=0$. As a consequence the energy per 
nucleon increases with increasing $g$. The EOS $\epsilon (p)$, on the other hand, turns out to be
independent of $g$. 

This picture holds qualitatively also for the case of SQM, where a non-vanishing
current mass of the strange quark is considered. Even in the case of a small bag constant
$B_0^{1/4}=145$ MeV and a small strange mass of $m_s=150$ MeV the energy per baryon $E/A$
of SQM is larger than the one of $^{56}F\!e$ already for $g=3$. For higher values of $m_s$,
$B_0^{1/4}$, or $g$ even larger values for $E/A$ are found.
This suggests that SQM is less
bound if medium effects are taken into account. However, it might still be metastable,
i.e. stable against decays caused by the strong interaction, which then has interesting
consequences for the formation of strangelets in ultrarelativistic heavy ion collisions
\cite{Greiner}.

Applying our results to strange quark stars, we observe that the radius-mass relation of
these stars is hardly changed by the presence of an effective quark mass since the EOS
depends only very weakly on it. However, owing to the increase of $E/A$ with an increasing
effective mass the phase transition to hadronic matter will take place at a smaller
radius in the interior of the star leading to a thicker crust.

\end{multicols}

\bigskip

\leftline{\bf Acknowledgements}

\medskip

We would like to thank N. Glendenning, A. Peshier and F. Weber for helpful discussions. K.S.
and M.H.T. are grateful to the University of Regensburg for their hospitality.



\begin{figure}[ht]
\centerline{\rotate[r]{\epsfig{file=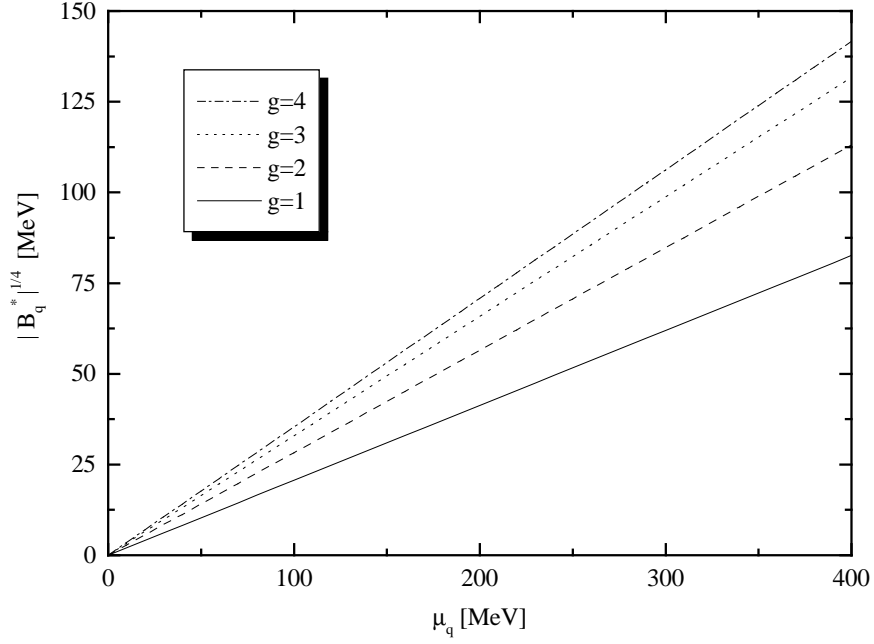,height=14.5cm}}}
\caption{$B_q^*(\mu)$ (\ref{Bq}) for various values of $g$; $d_q=18$ (assuming $m_s=0$)}
\label{bqmu}
\end{figure}

\end{document}